# Modeling Ferrimagnets in MuMax3: Temperature-Dependent Skyrmion Dynamics

*Valerii Antonov, Mikhail Letushev, Michail Bazrov, Zhimba Namsaraev, Ekaterina Steblii, Aleksey Kozlov, Aleksandr Davydenko, Maksim Stebliy*

*Laboratory of Spin–Orbitronics, Institute of High Technologies and Advanced Materials, Far Eastern Federal University, Vladivostok 690950, Russia*

Email: stebliyme@gmail.com

## Abstract

In this work, we propose an approach to modeling ferrimagnets in MuMax3. We show that by specifying two interacting magnetic sublattices as separate layers, it is possible to reproduce a sperimagnetic-like ordering of magnetization. In such a system, magnetic and angular momentum compensation states can be achieved only by varying the temperature while keeping other parameters fixed. This behavior arises from the different temperature dependencies of the magnetization projections in the sublattices, as determined by their thermal functions. We also investigated the motion of a skyrmion under the action of a spin-polarized current. By changing the temperature, we observed both the disappearance of the skyrmion Hall effect at the angular compensation point and the maximum velocity of translational motion. The latter effect requires a modified version of MuMax3 that allows the g-factor to be specified for different regions. The proposed approach can also be applied to study other phenomena in ferrimagnets, including the influence of composition on the magnetic and angular compensation temperatures, the tilted phase, domain-wall motion, and effects arising from non-uniform current or temperature distributions.

## 1. Introduction

Devices based on ferrimagnetic (FiM) materials have attracted significant attention for spintronic applications owing to their high stability, fast magnetization dynamics, and efficient current-induced manipulation [1] [2]. A ferrimagnet consists of two antiferromagnetically coupled sublattices: a transition-metal (TM) sublattice and a rare-earth-metal (RE) sublattice. Because these sublattices are inequivalent, ferrimagnets combine the usability of ferromagnets with advantages commonly associated with antiferromagnets, including exceptional stability and ultrafast dynamics. These benefits arise from unique ferrimagnetic states, such as the magnetization and angular momentum compensation points, where equal and opposite contributions from the sublattices cancel [3] [4]. In addition, ferrimagnets often exhibit practically important properties, including bulk perpendicular magnetic anisotropy (PMA) [5] [6], the Dzyaloshinskii–Moriya interaction (DMI) [7] [8], and the potential for internally generated effective magnetic fields under current flow, which can enhance spin–orbit torque efficiency [9] [10].

Over the past decade, research on ferrimagnetic systems has expanded rapidly, driven by advances in both experiment and theory. Among theoretical approaches, micromagnetic simulations have proven invaluable for understanding magnetization dynamics at the nanoscale. However, conventional micromagnetic software is often limited in its applicability to ferrimagnets. Accurate modeling typically requires atomistic methods, such as the VAMPIRE package [11] [12], which explicitly resolve the individual magnetic sublattices. While atomistic models provide high fidelity [13] [14], they are computationally intensive and generally impractical for simulating large-scale phenomena, including the dynamics of magnetic quasiparticles such as domain walls and skyrmions. More recently, codes such as MuMax+ [15] and BORIS [16] have introduced support for two-sublattice models and are likely to become widely adopted as they mature.

To address these challenges, researchers commonly employ effective macroscopic models of ferrimagnets within micromagnetic frameworks such as MuMax3 [17] or OOMMF [18], or solve coupled Landau–Lifshitz equations for two antiferromagnetically coupled sublattices [19] [20] [21]. Using these approaches, numerous static and dynamic properties of ferrimagnetic systems have been investigated, including domain-wall velocity [20] [22] [23], skyrmion dynamics [24] [19] [25], spin-wave propagation [26] [27] and thermal excitation [28]. Such studies typically aim to optimize simulation techniques for specific physical phenomena.

In this work, we develop and evaluate a broadly applicable equivalent model of ferrimagnets for use in MuMax3. In this model, a ferrimagnetic alloy is represented by two magnetic layers with specific interactions and material parameters. The approach reproduces relatively simple features, such as the temperature dependence of the net magnetization, as well as more complex phenomena, including the disappearance of skyrmion deflection during current-driven motion at the angular momentum compensation point.

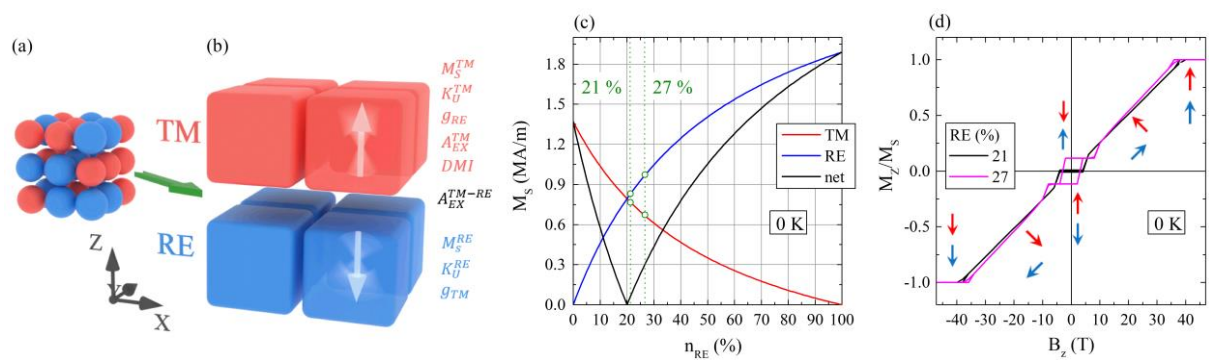

Fig.1. (a) An amorphous ferrimagnetic alloy (FiM) composed of a transition metal (TM) and a rare-earth metal (RE) is modeled as two separate layers of fixed thickness, each with its own set of parameters. (b) Saturation magnetizations of the TM and RE layers, as well as the net magnetization of the FiM at 0 K, as functions of RE content in the alloy. The green lines indicate the FiM parameters

*selected for plotting the hysteresis loops. The arrows schematically represent the magnetization orientations in the layers.*

## 2. Description of the equivalent model

Ferrimagnetic materials can be crystalline or, more commonly, amorphous, composed of alloys [3] [4] or multilayers [29] [30] [31] of transition-metal and rare-earth elements (Fig. 1a). In all cases, the magnetic structure consists of two coupled sublattices formed by TM and RE atoms. In the proposed model, these sublattices are represented as fixed-thickness layers (Fig. 1b), each characterized by distinct magnetic parameters: saturation magnetization (Ms), exchange stiffness (Aex), uniaxial anisotropy (Ku), Dzyaloshinskii–Moriya interaction (D), spin-current susceptibility (χ), damping constant (α), and gyromagnetic ratio (γ). The full set of parameters is listed in Table 1.

In our model, the total alloy volume (V) is fixed and partitioned into two layers (TM and RE). Variations in the RE concentration $n_{RE}$ are represented through coupled changes in the layers' saturation magnetizations ($M_{RE}^S$ and $M_{TM}^S$). We use a semi-analytical model, $M_S^{TM} = M_0^{TM} \frac{(1-n_{RE})}{(1-n_{RE})+n_{RE}k_v}$ and $M_S^{RE} = M_0^{TM} \frac{n_{RE}k_\mu}{(1-n_{RE})+n_{RE}k_v}$, where $M_0^{TM}$ is the saturation magnetization of the pure TM layer, $n_{RE}$ the fraction of RE atoms in the alloy, $k_\mu = \frac{\mu_{RE}}{\mu_{TM}}$ is the ratio of atomic magnetic moments, and $k_v = \frac{v_{RE}}{v_{TM}}$ is the ratio of atomic volumes. The model is parameterized using experimental data for CoTb alloys [32] [33]: the magnetic compensation state at $n_{RE} = 0.2$ suggests $k_\mu = 4$. Tabular values are used for the volume ratio $k_v = 2.9$ and $M_0^{TM} = 1.4 \times 10^6 A/m$. The resulting net magnetization, obtained by subtracting the two sublattice contributions, is shown in Fig. 1c.

The model assumes composition-independent atomic moments, which is a simplification since the moments vary significantly with composition and thickness [34] [35]. A nearest-neighbor model introduced in [34] accounts for the reduction of both RE and Co moments with increasing RE content. However, in this work only temperature effects were studied at a fixed composition near the compensation state, corresponding to RE dominance at 0 K.

In RE atoms, magnetic moments originate from the 4f shell: purely spin in Gd, and both spin and orbital in Tb [36] [37] [38]. Because the overlap of 4f orbitals is minimal, direct exchange interactions are extremely weak, and spontaneous ordering does not occur within the RE sublattice. Instead, TM atoms align the RE moments antiparallel to their own through complex indirect interactions [39]. Accordingly, the model includes a ferromagnetic exchange interaction between TM cells, an antiferromagnetic exchange interaction between TM and RE cells ($A_{ex}^{TM-RE}$), and no exchange interaction between RE cells.

The origin of perpendicular magnetic anisotropy in amorphous alloys remains not fully understood, but several mechanisms have been proposed [40] [41] [42] [43]: single-ion anisotropy, stress-induced (magnetostrictive) anisotropy, shape anisotropy arising from macroscopic structural inhomogeneities (e.g., columnar growth), and pair ordering due to non-uniform atomic distributions, where dipole–dipole interactions define a preferred axis. In Tb-containing alloys, single-ion anisotropy is typically regarded as dominant. Tb possesses a non-spherical 4f orbital that couples to the local crystal field and, through spin–orbit interaction, favors magnetization alignment perpendicular to the film plane. This anisotropy is transferred to the entire magnetic system via exchange with the TM sublattice, making the RE sublattice the primary source of anisotropy. In contrast, Gd-based systems also exhibit PMA despite Gd's spherical 4f orbital, implying that additional mechanisms contribute and that both sublattices may provide anisotropy. Given this uncertainty, our simulations assume equal anisotropy energy in both sublattices. Nevertheless, by adjusting other parameters, we were able to reproduce equivalent results under the assumption that anisotropy is present only in the RE layer.

Amorphous ferrimagnetic alloys such as CoTb or CoGd are characterized by the presence of Dzyaloshinskii–Moriya interaction (DMI). Its strength depends only weakly on the capping layer material [44] but strongly on alloy thickness [45], suggesting a bulk origin, presumably due to an asymmetric distribution of the elemental constituents. In the proposed model, DMI is assumed to arise primarily within the TM sublattice [7]. The 4f electrons in RE atoms, which generate the magnetic moment and exhibit strong spin–orbit coupling in the case of Tb, are deeply bound and highly localized. Because they lack significant orbital overlap with neighboring atoms, these 4f electrons cannot directly participate in the DMI mechanism.

Another important consequence of the localization of the RE 4f electrons, which are responsible for the magnetic moment, far below the Fermi level is that they do not contribute to magnetotransport effects [46]. This viewpoint is widely accepted, although some studies suggest a possible contribution from the RE sublattice [47] [48] [49]. Including such a contribution, however, greatly complicates the analysis without providing clear advantages. Therefore, in our model, only the interaction of the spin-polarized current with the TM layer was considered, while the RE layer was excluded from this interaction.

Tab.1.

| Parameter | TM | RE |
| --- | --- | --- |
| $M_{S\_0K}$ (MA/m) | 0.79 | 0.81 |
| $A_{ex}$ (pJ/m) | 12 | 0 |
| $A_{RE-TM}$ (pJ/m) | -8 | |
| $K_U$ (MJ/m$^3$) | 0.73 | 0.73 |
| $D_{DMI}$ (mJ/m$^2$) | 2.4 | 0 |
| $j_s$ (TA/m$^2$) | 0.2 | 0 |
| $\alpha$ | 0.019 | 0.019 |
| g | 2.1 | 2.0 |

## 3. Results

Using the proposed model, we show that a change in temperature alone, with all other parameters fixed, is sufficient to produce several key effects. First, the ferrimagnet crosses the magnetic compensation state. At the micromagnetic level, this arises from sperimagnetic-like ordering in the two sublattices, as described in Chapter 1. Second, a ferrimagnetic skyrmion reverses its deflection during current-induced motion. This reversal occurs when the system passes through the angular momentum compensation state, as discussed in Chapter 2.

### 3.1. Effect of temperature on the magnetization

In this work, the saturation magnetization was set to $M_S^{TM}(0) = 0.79\ MA/m$ for the TM layer and $M_S^{RE}(0) = 0.81\ MA/m$ for the RE layer, corresponding to the case of 21% RE atom content (Fig. 1c). At 0 K, the magnetic moments in both layers are strictly antiparallel and aligned with the anisotropy axis (Fig. 1d). Because the RE layer has the larger projection of its magnetic moment along the z-axis, the system is RE-dominant. The magnetization of the RE layer aligns with the external field, whereas the TM layer aligns in the opposite direction. Breaking the antiferromagnetic order requires an external field of a few tesla, while fields of several tens of tesla are needed to align the magnetizations of both layers in the same direction. The resulting hysteresis loop reproduces the qualitative features of the experimental data [40] [16] [50].

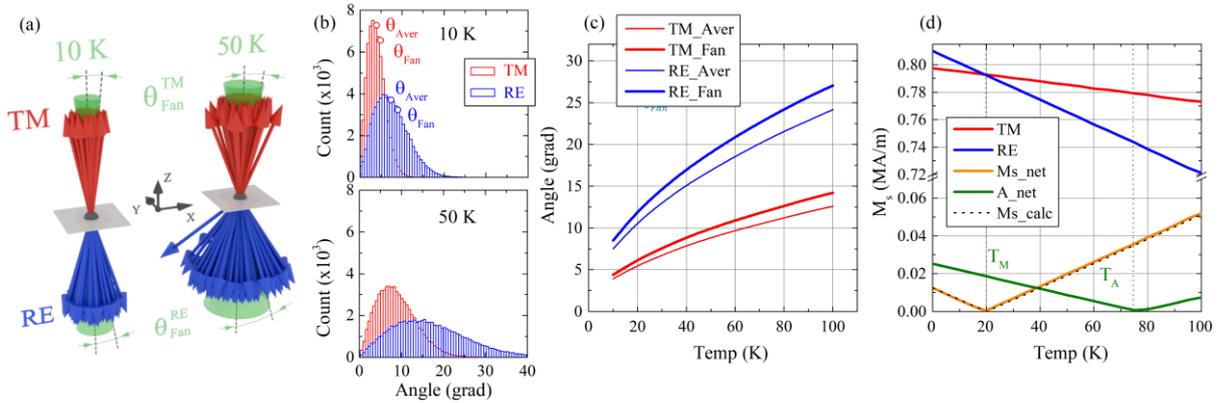

*Fig.2. (a) Examples of magnetic moment orientations in the TM and RE layers at 10 K and 50 K. Green cones represent the fanning angle that defines the resulting magnetization of each layer. (b) Distribution of deflection angles of magnetic moments in the layers at 10 K and 50 K, obtained from a 256×256 cell simulation. (c) Temperature dependence of the average deflection angle and fanning angle for each layer. (d) Resulting magnetizations of the TM and RE layers, and the net magnetization of the FiM, as functions of temperature. The black line shows the result of analytical fitting, while the green line represents the calculated net angular momentum.*

As the temperature increases to 10 K, for example, magnetic moments deviate from the anisotropy axis due to thermal fluctuations, producing a spread of orientations in both layers (Fig. 2a). The resulting non-Gaussian distributions of deviation angles are described in terms of the polar angle θ. The peak marks the most probable angle, whereas the average angle $\theta_{Ave}$ lies to the right of it. To determine the z-axis projection of the magnetization in a layer, the average value of cosine ($\langle cos\theta \rangle$) should be used rather than the cosine of the average angle ($cos\theta_{Ave}$). The corresponding angle $\theta_{Fan}$ describes the fanning cone (Fig. 2a). The angles $\theta_{Ave}$ and $\theta_{Fan}$ are indicated on the distributions for the TM and RE sublattices (Fig. 2b). At the same temperature, the RE sublattice exhibits greater dispersion and larger deviation angles than the TM sublattice, indicating stronger fluctuations and higher thermal mobility of the RE magnetic moments, which result from the absence of direct exchange interactions within the RE sublattice.

With further increase in temperature to 50 K, the spread of orientations grows (Fig. 2a). As shown in Fig. 2c, the fanning angle of the sublattices changes with temperature at different rates, increasing more rapidly for the RE sublattice. Consequently, at 50 K, the projection of the RE sublattice magnetization onto the z-axis becomes smaller than that of the TM sublattice, and the ferrimagnet becomes TM-dominant. Since the system is RE-dominant at 0 and 10 K, it passes through a magnetic compensation state when all parameters except temperature are fixed. This behavior qualitatively corresponds to the sperimagnetic-like ordering used to describe ferrimagnets [1] [40] [51] and arises naturally in the present model.

Figure 2d shows the temperature dependence of the magnetization of the TM and RE layers for the parameters listed in Table 1, excluding current effects, which are not considered here (the corresponding simulation file is provided in the supplementary materials, *File_1*). The magnetization of each layer was calculated as the sum of all magnetic moment projections onto the anisotropy axis (z-axis). The results follow the analytical expressions given by $M_S^{TM}(T) = M_S^{TM}(0)(1 - T/T_C)^{\beta_{TM}}$ and $M_S^{RE}(T) = M_S^{RE}(0)(1 - T/T_C)^{\beta_{RE}}$, where $M_S(0)$ denotes the saturation magnetization at zero temperature, $T_C$ is the Curie temperature, and β is the critical exponent. [52]. The net magnetization of the system, obtained as the difference between the two sublattice contributions ($|M_{net}| = |M_{TM}| - |M_{RE}|$), agrees closely with the analytical model. The data reveal a transition through the magnetic compensation state at $T_M = 20\ K$.

Another important state of a ferrimagnetic system is the angular momentum compensation state, which can be calculated as $|A_{net}| = \left|\frac{M_S^{TM}}{\gamma_{TM}}\right| - \left|\frac{M_S^{RE}}{\gamma_{RE}}\right|$, where $\gamma_{TM} = g_{TM}\frac{\mu_B}{\hbar}$ and $\gamma_{RE} = g_{RE}\frac{\mu_B}{\hbar}$ are the gyromagnetic ratios of the TM and RE sublattices, $\mu_B$ is the Bohr magneton, and $\hbar$ is the reduced Planck constant [1] [53]. Using $M_S^{TM}$ and $M_S^{RE}$ from the simulated dependences, together with g-factors g1 and g2 from Table 1, the temperature dependence of the angular momentum was obtained and is shown in Fig. 2d, yielding $T_A = 75\ K$. Additional simulations confirmed that the magnetic compensation temperature $T_M$ does not depend on the g-factors.

The system parameters in Table 1 were selected so that the magnetic compensation temperature is 20 K. This choice allows the study of skyrmions to be restricted to temperatures up to 100 K, since at higher temperatures thermal fluctuations strongly distort their shape. However, by adjusting the initial magnetizations of the TM and RE layers, as well as the anisotropy and exchange energies within reasonable limits, any desired magnetic compensation temperature can be achieved.

### 3.2. Effect of temperature on the skyrmion dynamics

Using the parameters from Table 1, a Néel skyrmion was initialized as the magnetic configuration and remained stable in the absence of external fields throughout the investigated temperature range. Skyrmion motion was then studied under the action of a spin current applied along the +z direction with -y polarization (Fig. 3a). In the model, the current was injected only into the TM layer, where it induced magnetization changes via the spin–orbit torque mechanism [54]. The corresponding simulation script is provided in the supplementary materials (File_2).

The action of the spin-polarized current on the TM layer drives the skyrmion in the +x direction. At $T_1 = 0$ K, in which the ferrimagnet is RE-dominant, the skyrmion is deflected toward the +y direction (Fig.3b at $T_1$). The motion can be qualitatively described by the Thiele equation ($\boldsymbol{G} \times \boldsymbol{v} - \alpha \boldsymbol{D}\boldsymbol{v} + \boldsymbol{F}_{SOT} = 0$), where $\boldsymbol{G} = G\hat{z}$ is the gyromagnetic vector (proportional to the net angular momentum and topological charge), $\boldsymbol{v} = v_x\hat{x} + v_y\hat{x}$ is the skyrmion velocity, $\alpha$ is the damping parameter $\boldsymbol{D}$, is the damping tensor, $\boldsymbol{F}_{SOT}$ is the force due to spin-orbit torque [55] [56]. The cross-product term ($\boldsymbol{G} \times \boldsymbol{v}$) produces an additional Magnus force perpendicular to the skyrmion velocity, resulting in transverse deflection—the skyrmion Hall effect [57] (Fig. 1b)

As the temperature increases to a certain value $T_2 > T_1$, the skyrmion moves in a straight line without deflection (Fig. 3a). A further increase to $T_3 > T_2$ causes the skyrmion to deviate in the opposite direction (-y) during motion. This reversal of the deflection direction indicates a transition through the angular momentum compensation state at temperature $T_A$. At this point, the gyrovector vanishes, and the skyrmion Hall effect disappears [58] [59] [24]. Quantitatively, $T_A$ can be determined from the temperature dependence of the skyrmion velocity along the y-axis. As shown in the Fig. 3c, $v_y$ changes sign and passes through zero, corresponding to the angular momentum compensation point.

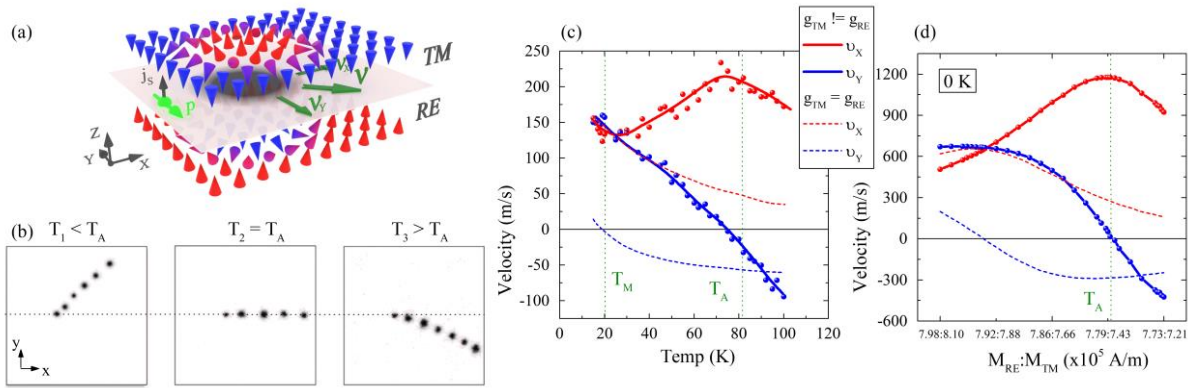

Fig.3. (a) Diagram of a skyrmion in the double-layer FiM. Arrows indicate the directions of the velocity components $\vartheta_X$ and $\vartheta_Y$, current density $j_S$, and spin polarization $p$. (b) Skyrmion motion trajectories under the influence of spin current at temperatures $T_1$ (below the angular momentum compensation

*temperature $T_2$) and $T_3$ (above $T_2$). (c) Temperature dependence of the skyrmion velocity components for the cases with equal and different $g-factors$ in the layers. (d) Dependence of the skyrmion velocity components at $0\ K$ on the saturation magnetization of the RE and TM layers, with values corresponding to the different temperatures shown in Fig. 2d.*

The temperatures $T_1$, $T_2$, and $T_3$ in Fig. 3b depend on the g-factors of the layers. It should be noted that the current version of MuMax3 does not allow different g-factor values to be assigned to different regions. Therefore, in the standard unmodified version of the program, the same value $g_{TM} = g_{RE} = 2$ was used. In this case, the angular momentum compensation temperature coincides with the magnetic moment compensation temperature ($T_A = T_M = 20\ K$).

Since MuMax3 is open source, user-modified versions are available. The version kindly published by João Sampaio allows the g-factor to be assigned separately for each region [60]. Using $g_{TM} = 2.1$ and $g_{RE} = 2.0$ in this version produces a separation between the compensation temperatures: $T_M = 20\ K$ and $T_A = 75\ K$ (Fig. 3b). The simulations also reveal an important effect: the appearance of a maximum in skyrmion velocity at the angular momentum compensation temperature when different g-factors are used (Fig. 1). This effect has been predicted theoretically and observed experimentally in studies of domain-wall and skyrmion velocities [24] [61] [62] [63].

## 4. Discussion

The temperature dependence of the net magnetization in a ferrimagnet can be explained by the reduced projection of the magnetization onto the anisotropy axis in relatively large cells, while the magnetic moment of each individual cell remains unchanged. This description is statistical, but the qualitative changes in skyrmion motion demonstrate that these effects can be represented as changes in the effective saturation magnetization of the sublattices. As shown in Fig. 1d, the angular momentum compensation temperature obtained from the calculations coincides with the temperature corresponding to the maximum skyrmion velocity (Fig. 3c).

The qualitative explanation for the maximum skyrmion velocity can be attributed to several factors. At the angular momentum compensation point, the gyrovector G approaches zero, causing the skyrmion Hall angle to vanish and eliminating transverse deflection. As a result, the effective force produced by spin–orbit torque is directed almost entirely along the forward direction, maximizing the longitudinal velocity [24] [59]. However, the most notable feature Fof angular momentum compensation, however, is the acceleration of magnetic dynamics. The micromagnetic behavior of a ferrimagnet can be understood as a superposition of two collective precession modes: a ferromagnetic mode, where the sublattice magnetizations rotate in phase and determine the orientation of the net magnetization, and an exchange mode, where the sublattices precess in antiphase. Near the angular momentum compensation point, the ferromagnetic mode frequency rises sharply while the exchange mode frequency decreases, leading to faster dynamics and enhanced responsiveness to external stimuli [64] [65] [66]. This effect may also be linked to a reduction in the effective skyrmion mass [67] [68], although a rigorous theoretical description remains unavailable.

For comprehensive validation, skyrmion motion was also simulated at 0 K using the layer saturation magnetizations $M_S^{TM}$ and $M_S^{RE}$ obtained from the temperature dependence in Fig. 2d. In this approach, the magnetic moment of each cell is specified explicitly. Under these conditions, both the maximum velocity and the disappearance of the skyrmion Hall effect are observed at magnetization values corresponding to temperature $T_A$ (Fig. 2d). A notable difference, however, is the magnitude of the velocity change: the difference reaches a factor of five, compared to only about a factor of two in the temperature-driven case. The reduction in velocity with increasing temperature is attributed to two effects, clearly visible in Fig. 1: (1) deformation of the skyrmion shape, which enhances dissipation, and (2) thermal fluctuations in the medium, which provide an additional dissipation channel.

Thus, the proposed model reproduces both magnetic and angular momentum compensation as a function of temperature, as well as a more complex effect: the change in skyrmion deflection under spin current and the emergence of a velocity maximum. The same approach may also be applied to analyze other ferrimagnetic phenomena, including (1) the dependence of magnetic and angular momentum compensation temperatures on alloy composition, (2) the appearance of a tilted phase associated with the breakdown of antiferromagnetic ordering between sublattices, (3) the dynamics of domain-wall motion, and (4) the influence of non-uniformity.

## 5. Conclusion

In this work, the approach to modeling ferrimagnets in MuMax3 was developed and tested. The amorphous ferrimagnetic alloy was represented as two antiferromagnetically coupled layers of fixed thickness, each with distinct interactions. This model naturally reproduces sperimagnetic ordering and enables the realization of magnetic compensation by varying temperature. It also allows the stabilization of skyrmions and the study of their dynamics under spin-polarized current as a function of temperature. The simulations demonstrate that at the angular momentum compensation temperature, the skyrmion Hall angle vanishes and the skyrmion velocity reaches a maximum. The latter effect is observed when the two magnetic layers are assigned different gyromagnetic ratios.

## 6. Acknowledgments

We thank Dr. João Sampaio for insightful consultations that contributed to this work.

Micromagnetic modeling was supported by the Russian Science Foundation, grant No. 25-42-00083, https://rscf.ru/project/ 25-42-00083. Programming of the code for simulation of multilayer systems with different g-factors in layers was financed by the Russian Ministry of Science and Higher Education (State Assignment No. FZNS–2023–0012). Analytical calculations were funded by the Russian Science Foundation, grant No. 23-42-00076, https://rscf.ru/project/23-42-00076.

FiM_Magnetization_on_temperature.mx3
```
/***************************************************************
  Description:
  This mumax3 scripts determines the total projection of the magnetization of
  each of the two layers onto the z-axis, depending on the temperature.

***************************************************************/

    setgridsize(256, 256, 2)
    setcellsize(1e-9, 1e-9, 1e-9)
    setpbc(4, 4, 0)
    OutputFormat = OVF2_TEXT

    // Define two regions along the z-axis (two layers)
    defregion(1, zrange(-1e-9,0e-9))  // TM layer
    defregion(2, zrange(0e-9, 1e-9))  // RE layer

    // Set material parameters for TM layer
    Aex.setregion(1, 12e-12)
    Dind.setregion(1, 2.4e-3)
    Msat.setregion(1, 0.78e6)
    Ku1.setregion(1, 0.73e6)
    anisU.setRegion(1, vector(0, 0, 1))

    // Set material parameters for RE layer
    Aex.setregion(2, 0e-12)
    Dind.setregion(2, 0e-3)
    Msat.setregion(2, 0.88e6)
    Ku1.setregion(2, 0.73e6)
    anisU.setRegion(2, vector(0, 0, 1))

    // Set interlayer exchange coupling between TM and RE regions
    ext_InterExchange(1, 2, -8e-12)

    // Configure data output tables
    tableAdd(Temp)             // Record temperature
    TableAdd(m_full.region(1)) // Record magnetization of TM region
    TableAdd(m_full.region(2)) // Record magnetization of RE region
    TableAdd(m_full)           // Record total magnetization

    // Set physical constants and damping parameters
    gammaLL=1.716e11           // Gyromagnetic ratio (rad/s·T)
    GFactor.setregion(1,2.1)   // g-factor for region 1
    GFactor.setregion(2,2.0)   // g-factor for region 2
    alpha.setregion(1,0.019)   // Gilbert damping for region 1
    alpha.setregion(2,0.019)   // Gilbert damping for region 2

    // Set initial magnetization states
    m.setRegion(1, uniform(0,0,-1))
    m.setRegion(2, uniform(0,0,1))

    // Main simulation loop: temperature sweep from 0K to 190K in 10K steps
    for i:=0; i<20; i++ {
       Temp = i*10  // Set current temperature (0, 10, 20, ..., 190K)
       run(1e-9)
       tablesave()
    }
```

Velocity_FiM_skyrmion.mx3
/*****************************************************************

Description:
The program script simulates the motion of a skyrmion in a two-layer system
under the influence of a current. The skyrmion velocity at each temperature
is determined by analyzing the dependence of the skyrmion coordinate on time.

Using different g-factors across layers requires using a modified version of MuMax3:
https://github.com/mumax/3/pull/315

This modification introduces a new parameter, Gfactor, and allows it to be defined
for each region. This partitioning is taken into account in the functions responsible for
the Zhang-Li torque. However, skyrmion motion in SOT structures is primarily driven by
the Slonczewski-like torque. Therefore, for the model to work correctly, additional
modifications are required to include Gfactor as an argument for the functions that
define the Slonczewski-like torque in Mumax3.

*****************************************************************/

```
setgridsize(256, 256, 2)
setcellsize(1e-9, 1e-9, 1e-9)
setpbc(4, 4, 0)
OutputFormat = OVF2_TEXT

// SOT (Spin-Orbit Torque) parameters for current-induced magnetization dynamics
xi = 0.1       // non-adiabaticity of spin-transfer-torque
Pol = 1        // Polarization efficiency
fixedlayer = vector(0, -1, 0)  // Fixed layer direction for spin torque reference

// Define two magnetic layers along the z-axis
defregion(1, zrange(-1e-9,0e-9))  // TM layer
defregion(2, zrange(0e-9, 1e-9))  // RE layer

Msat.setregion(1, 0.7975e6)
Msat.setregion(2, 0.81e6)
Aex.setregion(1, 12e-12)
Aex.setregion(2, 0e-12)
Dind.setregion(1, 2.4e-3)
Dind.setregion(2, 0e-3)
Ku1.setregion(1, 0.73e6)
anisU.setRegion(1, vector(0, 0, 1))
Ku1.setregion(2, 0.73e6)
anisU.setRegion(2, vector(0, 0, 1))

// Interlayer exchange coupling between the two regions
ext_InterExchange(1, 2, -8e-12)

// Physical constants for LLG equation
alpha.setregion(1,0.019)   // Gilbert damping parameter for the TM region
alpha.setregion(2,0.019)   // Gilbert damping parameter for the RE region

// Only valid for the modified version of the program
//gammaLL=1.716e11          // Gyromagnetic ratio (rad/s·T)
//GFactor.setregion(1,2.1)  // Landé g-factor for the TM region
//GFactor.setregion(2,2.0)  // Landé g-factor for the RE region

// Initial magnetization configuration: coupled antiferromagnetic skyrmions
m.setRegion(1, NeelSkyrmion(-1,1).transl(-40e-9,0,0))  // TM Region: skyrmion with core down
m.setRegion(2, NeelSkyrmion(1,-1).transl(-40e-9,0,0))  // RE Region: skyrmion with core up

// Simulation conditions
Temp=80  // Temperature set to 80K (affects thermal fluctuations)

// Electric current density applied to region 1 only
J.setregion(1, vector(0,0,0.2e12))  // Current density 0.2×10¹² A/m² along z-direction

// Configure data collection for analysis
TableAdd(m_full.region(1))   // Record magnetization of the TM layer
TableAdd(m_full.region(2))   // Record magnetization of the RE layer
tableAdd(ext_bubblepos)      // Record skyrmion position
```

```
// Configure automatic saving during simulation
autosave(m,1e-10)      // Save magnetization configuration every 0.1 ns
tableAutosave(1e-11)   // Save table data every 0.01 ns

// Run the simulation for 1 nanosecond
run(10e-9)
```